\documentclass[aip,apl,reprint,superscriptaddress,amsmath,amssymb,showkeys,a4paper]{revtex4-1}

\usepackage[utf8]{inputenc}
\usepackage{setspace}
\usepackage{graphicx}
\usepackage{dcolumn}
\usepackage{bm}
\usepackage[T1]{fontenc}  
\usepackage{lmodern}   
\usepackage{subfigure}

\begin{document}

\title{Gate Bias Stress in Pentacene Field-Effect-Transistors: Charge Trapping in the Dielectric or Semiconductor}
 
 \author{R. Häusermann}
  \email{hroger@phys.ethz.ch}
  \author{B. Batlogg}
 \affiliation{ Laboratory for Solid State Physics, ETH Zurich, 8093 Zurich, Switzerland}

\date{\today}

\begin{abstract}
Gate bias stress instability in organic field-effect transistors (OFETs) is a major conceptual and device issue. This effect manifests itself by an undesirable shift of the transfer characteristics and is associated with long term charge trapping. We study the role of the dielectric and the semiconductor separately by producing OFETs with the same semiconductor (pentacene) combined with different dielectrics (SiO$_2$ and Cytop). We show, it is possible to fabricate devices which are immune to gate bias stress. For other material combinations, charge trapping occurs in the semiconductor alone, or in the dielectric.
\end{abstract}
\keywords{Gate Bias Stress, trapping, pentacene, Cytop, current instability, degradation, interface, threshold voltage shift}
\maketitle


Charge trapping in organic field-effect transistors (OFETs) happens on various time scales. Trapping due to gate bias stress is a relatively slow process which takes place on the timescale of fractions of seconds up to several days, posing a challenge to the experiment as well as the conceptual understanding. Previous studies produced OFETs with a gate bias stress stability comparable to that of a-Si thin-film-transistors (TFT) \cite{Umeda2008}.\\
Several mechanisms have been proposed to explain the threshold-voltage shift under gate bias stress: (i) trapping of charges in the bulk of the semiconductor \cite{Chang2006}, (ii) trapping in disordered areas of the semiconductor\cite{Salleo2005}, (iii) trapping in regions in-between crystalline grains of the semiconductor\cite{Tello2008}, (iv) trapping in states at the semiconductor/dielectric interface\cite{Street2006} and (v) formation of bipolarons in the semiconductor\cite{Street2003,Paasch2007} as summarized by Sharma et al. \cite{Sharma2010a}. From temperature dependent measurements gate bias stress effects were found to be thermally activated with a comparable activation energy of $E_{\tau}\approx 0.6$~eV over a wide range of organic materials\cite{Mathijssen2007}. This has been taken as an indication that the underlying process is the same for all tested organic materials. Recently, charges have been shown to be trapped in the SiO$_2$ as the gate dielectric \cite{Mathijssen2010}. Although the mechanism behind this trapping is still unclear, Sharma et al.\cite{Sharma2010,Sharma2010a} explain it by a proton migration mechanism involving water, whereas Lee et al. \cite{Lee2010c} assume direct drift/diffusion of holes into the dielectric.\\
Here we present a comparative study of OFETs with either SiO$_2$ or Cytop (a highly hydrophobic fluoropolymer) as dielectric and with pentacene thin-films or single crystals as semiconductor. The focus is on a quantitative measurement of the threshold voltage shift $\Delta$V$_{th}$ over an extended time range (10$^{-1}$ to 10$^{3}$~s) in response to an applied gate voltage. By combining the semiconductor either in thin-film or single crystal form  with the two different gate dielectrics, charge trapping can either be completely suppressed in the OFET, or it occurs (a) in the semiconductor or (b) in both the semiconductor and the dielectric.



 \begin{figure}
\centering
\includegraphics[width=\columnwidth]{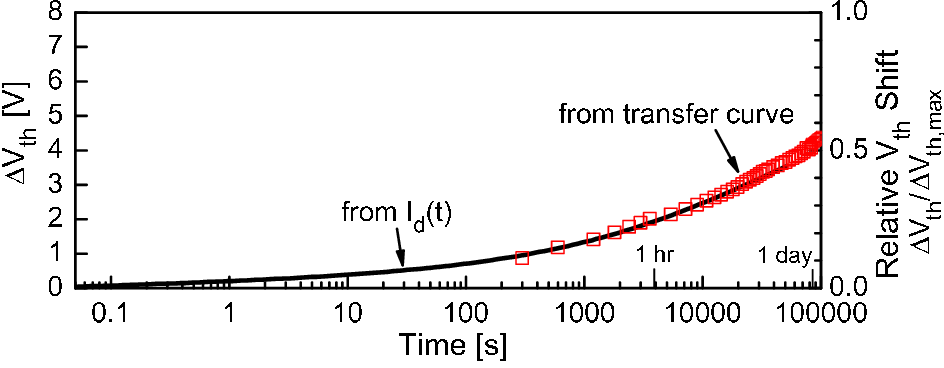}
\caption{ \label{fig:comparison_deltaVth} The influence of the Gate Bias Stress on the threshold Voltage V$_{th}$ has been measured in two ways: (i) the decrease of I$_d$(t) over time has been measured, (ii) V$_{th}$ has been calculated from the shift of the transfer curve over time. These two methods agree with each other in determining the threshold voltage shift $\Delta$V$_{th}$.}
\end{figure}

\begin{figure*}
\centering
{\includegraphics[width=0.9\textwidth]{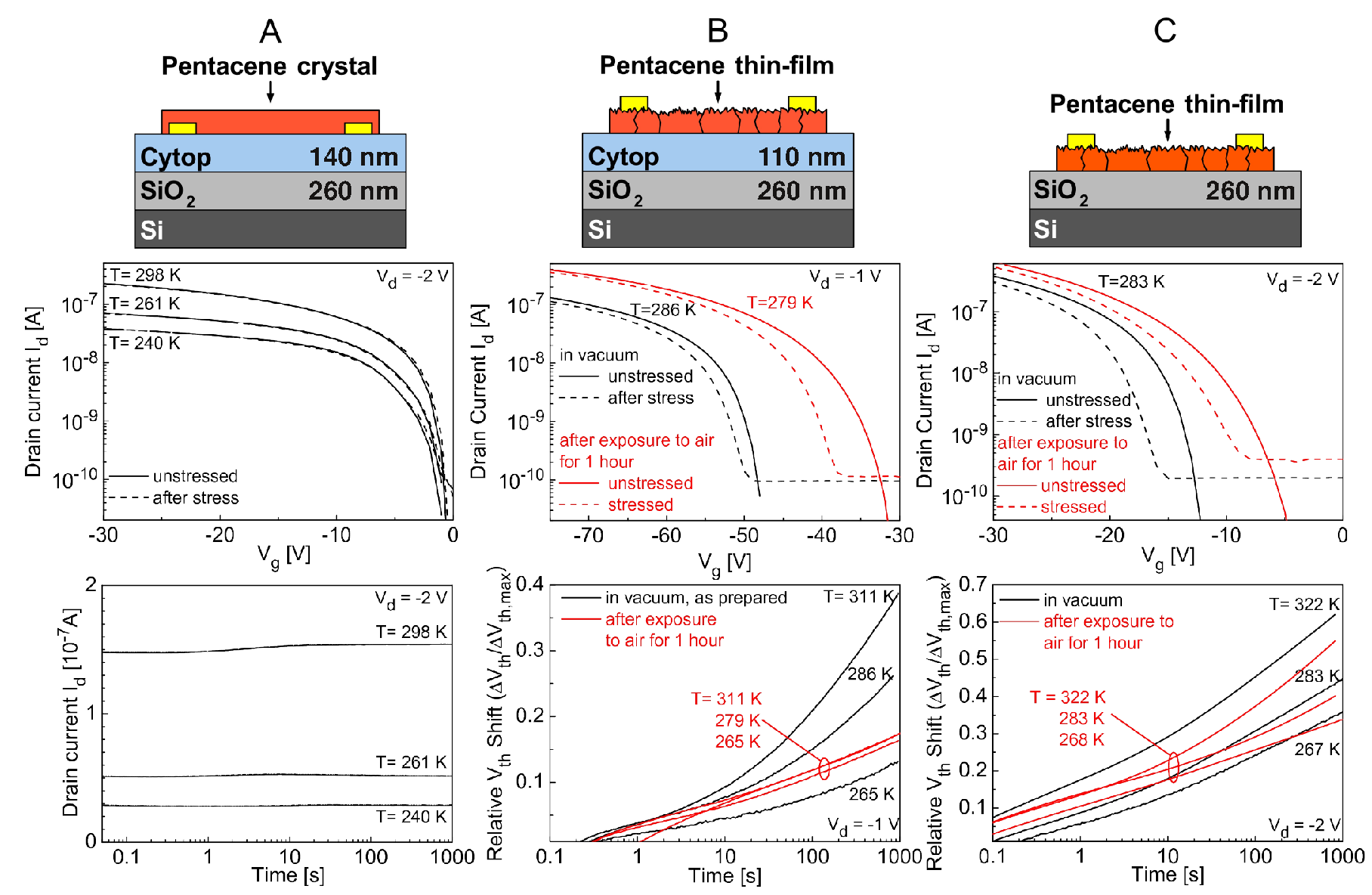}}
\caption{\label{fig:transfer_current_comp} Different mechanism at work resulting in gate bias stress effects. This compilation of graphs show the device setup (first row), the measured transfer curves (middle row) and the drain current I$_d$(t) or relative threshold voltage shift over time (bottom row).\\
\textbf{Device A:} (C$_i$=6.6$\times$10$^{-5}$~F/m$^2$) Transfer curves at various temperatures before and after stress do not show any gate bias stress effects. The same for the measurement of the drain current. $I_d(t)$ is stable over the measured time with a slight increase which is attributed to self heating of the device.\\
\textbf{Device B:} (C$_i$=7.4$\times$10$^{-5}$~F/m$^2$) Shows a gate bias stress effect in the transfer curves. The time and temperature dependent measurement of the relative threshold voltage shift (bottom row) shows this effect to strongly depend on temperature. After exposure to ambient air for 1 hour, the underlying process for long term charge trapping completely changes and becomes temperature independent. \\
\textbf{Device C:} (C$_i$=1.3$\times$10$^{-4}$~F/m$^2$) Also shows gate bias stress visible in the transfer curve and the time/temperature dependent measurement. After exposure to air, the the gate bias stress effect stays strongly temperature dependent.\\
}
\end{figure*}

Usually the threshold voltage shift $\Delta$V$_{th}$(t) is analyzed by measuring transfer curves at various times. With a gate bias applied in between. This straightforward measurement has some drawbacks, however: Firstly, it is slow, therefore information about the first few, or even fraction of seconds is not accessible. Secondly, some devices, especially thin-film-transistors (TFTs) do not show a linear regime and thus deriving a threshold voltage is not straightforward, even as the transfer curve clearly shifts. \\
Therefore we use a different method to measure the gate bias stress influence over time. Instead of measuring a full transfer curve, the evolution of the drain current I$_d$ is measured at a certain applied gate voltage V$_g$ and source-drain voltage V$_d$ \cite{Shih2007}. The decrease of I$_d(t)$ over time can be translated into a shift of the transfer curve (and therefore V$_{th}$). These two methods give the same result (Fig. \ref{fig:comparison_deltaVth}). A ``relative threshold voltage shift'' $\frac{\Delta V_{th}}{\Delta V_{th,max}}$ is used to compare different devices. $\Delta V_{th,max}$ is given by the difference between the initial threshold voltage $V_{th}$(t=0) and the applied gate bias $V_{g,appl}$.\\
The 3 devices, differing in their semiconductor/dielectric combination, are shown in figure \ref{fig:transfer_current_comp}. Device A is a flip-crystal OFET \cite{DeBoer2004,Takeya2003a} with Cytop as the gate dielectric \cite{Kalb2007,Walser2009} and the crystals grown according to Kloc et al. \cite{Kloc1997}. For Devices B and C, the pentacene thin-films and the gold contacts have been evaporated in 10$^{-7}$~mbar vacuum, the devices are transferred to the attached probe station, where they can be exposed to various atmospheres in a controlled way: in this study 1 hour of air at 0.2~bar was chosen. The single crystal device has been exposed to ambient air during sample preparation.\\
The devices have been analyzed as follows: A transfer-curve has been measured using a series of short V$_g$ pulses (50~ms) in order to minimize the stress. Afterwards the source-drain current, I$_d$(t), has been measured at a fixed V$_g$ and V$_{d}$. At the end, again a pulsed transfer-curve has been measured, with a V$_g$ offset between two pulses equal to the previously applied gate bias during stressing, to check that the shape of the transfer curve has been conserved. We repeat these series of measurements at several temperatures. In addition the transfer characteristics are the basis for calculating the trap density-of-states (trap DOS) in the band-gap using a FET simulator developed by Oberhoff et al. \cite{Oberhoff2007},  applied to other OFET configurations previously \cite{Kalb2010,Kalb2010b,Pernstich2006}.

\begin{figure}
\includegraphics[width=0.8\columnwidth]{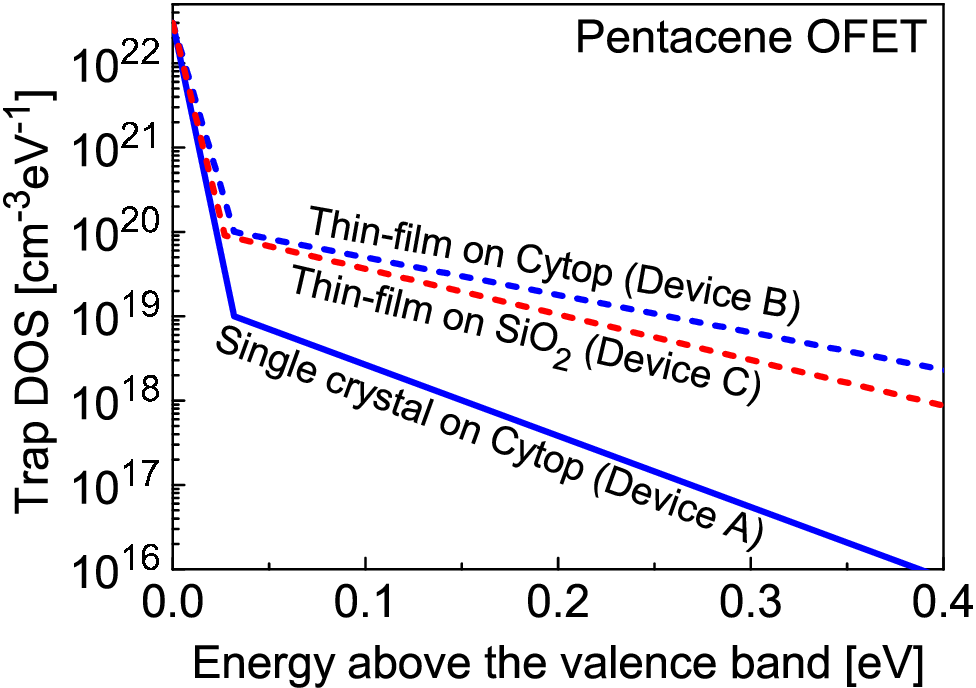}
\caption{ \label{fig:DOS_comparison} Comparison of different trap density-of-states for the 3 devices as calculated from transfer curves taken at different temperatures. The two thin-film devices have essentially the same trap DOS, whereas in the single crystal OFET the trap DOS is one to two orders of magnitude lower.}
\end{figure}

For Device A, the single crystal OFET, most significantly no gate bias stress effect is visible after stressing for 17 minutes at V$_g$=-20~V. Even when monitoring the source-drain current during applied gate voltage (Fig. \ref{fig:transfer_current_comp} bottom) no reduction of the current is seen which would indicate stress effects. The hole mobility is $\thickapprox$ 0.6 - 0.7~cm$^2$/Vs at room temperature. Additionally the trap DOS is very low (Fig. \ref{fig:DOS_comparison}). Therefore, the combination of Cytop together with a pentacene single-crystal results in an extremely good interface \cite{Kalb2007,Uno2008}, and long term charge trapping is effectively suppressed in the dielectric. Obviously, no long term charge trapping occurs in the pentacene crystal either.   

Device B, consisting of polycrystalline thin-film evaporated onto Cytop dielectric, shows gate bias stress effect. The extensive study of $\Delta$V$_{th}$(t) reveals long term charge trapping to be strongly temperature dependent when the device is always kept in vacuum. From Device A we learned that charges are not trapped in the Cytop, and thus in Device B it occurs in the polycrystalline pentacene thin-film. Exposing Device B to air causes several changes. The mobility approximately doubles from 0.08 to 0.15~cm$^2$/Vs. Turn-on and threshold voltage V$_{th}$(0) are significantly reduced. A more peculiar effect after exposure to air is seen when analyzing the temperature dependence of $\Delta$V$_{th}$(t): the gate bias stress effect becomes essentially independent of temperature.

Device C shows  similar turn-on behavior as Device B except at a much lower initial threshold voltage. Again $\Delta$V$_{th}$ due to gate bias stress is clearly visible in the transfer curves as well as in the threshold voltage shift over time. In passing, we note $\Delta$V$_{th}$(t) to be stronger by a factor of 2 in Device C compared to Device B. Exposure to air increases the mobility (0.1 to 0.15~cm$^2$/Vs) and shifts the transfer curves to lower gate voltage. After exposure to air, the relative threshold voltage shift remains temperature dependent, in contrast to Device B, where the temperature dependence vanishes after exposure to air. 

For further discussion, we would like to point out that the mobilities in the two thin-film devices are essentially identical and the trap DOS is similar too (Fig. \ref{fig:DOS_comparison}). Thus from the conduction measurements before and after exposure to air these two films behave the same. Therefore it appears that whatever their morphological differences may be, they have no effect on the DC properties. Additionally, as Miyadera et al. \cite{Miyadera2008} have found, the morphology of a pentacene film does not significantly change the long term trapping dynamics, thus the distinctly different behavior between Device B and C can not be attributed to a different film morphology.

The shift of the initial threshold voltage after exposure to air in Devices B and C is due to filling of charge traps, where oxygen acts as a hole donor. This shift is the same for Device B and C, in terms of induced charge (5$\times$10$^{18}$~e/cm$^{3}$), assuming a channel thickness of 1.5~nm. This is in quantitative agreement with the analysis by Kalb et al. \cite{Kalb2008}, where trap states were created in the band gap due to oxygen exposure. Therefore, since long term charge trapping in devices with SiO$_2$ is known to be due to trapping in the SiO$_2$ \cite{Mathijssen2010}, exposure to oxygen which alters the semiconductor only, does not change this mechanism significantly. But in devices where long term charge trapping occurs in the semiconductor, changes of the semiconductor due to oxygen exposure can be expected to alter this mechanism. This is exactly what has been observed as the difference between Device B and C.


In Summary we have shown that it is possible to fabricate OFETs using Cytop and pentacene single crystals which do not show gate bias stress effects and have a very low trap density. The device with pentacene thin-films on Cytop does show gate bias stress effects which is due to trapping of charges in the semiconductor. When this device is exposed to air, the gate bias stress effect becomes temperature independent which is an additional hint of the different origin of the charge trapping mechanism when compared to the OFET with SiO$_2$ as gate dielectric. Therefore we conclude that long term charge trapping occurs in the semiconductor thin-film if charge trapping in the dielectric is suppressed.

\end{document}